\documentclass{iaus}
\usepackage{graphicx}

\title[3-D hydrodynamic simulations of convection in A stars]
{3-D hydrodynamic simulations of \\convection in A stars}

\author[Kochukhov et al.]
{O. Kochukhov$^1$, B. Freytag$^1$, N. Piskunov$^1$ \and M. Steffen$^2$}

\affiliation{$^1$Astronomical Observatory, Uppsala
University, Box 515, 751 20, Uppsala, Sweden\\[\affilskip]
$^2$Astrophysikalisches Institut Potsdam, an der Sternwarte 16, 
14482 Potsdam, Germany}

\pubyear{2007}
\volume{IAUS239} 
\pagerange{1--3}
\date{?? and in revised form ??}
\jname{Convection in Astrophysics}
\editors{F. Kupka, I.W. Roxburgh \& K.L. Chan, eds.}

\newcommand{\cob}{CO$^5$BOLD}
\newcommand{\kms}{km\,s$^{-1}$}

\begin{document}

\maketitle

\begin{abstract}
Broadening and asymmetry of spectral lines in slowly rotating late A-type
stars provide evidence for high-amplitude convective motions. The properties
of turbulence observed in the A-star atmospheres are not understood theoretically and 
contradict results of previous numerical simulations of convection. Here we describe 
an ongoing effort to understand the puzzling convection signatures of A stars 
with the help of 3-D hydrodynamic simulations. Our approach combines realistic 
spectrum synthesis and non-grey hydrodynamic models computed with the \cob\ code. 
We discuss these theoretical predictions and confront them with high-resolution
spectra of A stars. Our models have, for the first time, succeeded in reproducing the 
observed profiles of weak spectral lines without introducing fudge broadening parameters. 
\end{abstract}

\firstsection 
              
\section{Introduction}

A number of complex hydrodynamic processes takes place in the atmospheres of cool A stars 
($T_{\rm eff}$\,=\,7500--8000~K). Convection is arguably the least understood phenomenon,  as
it presents enormous challenges both for observational detection and theoretical modelling. The
atmospheric structure of cool A stars is thought to be close to radiative equilibrium.
Nevertheless,  prominent signatures of convective motions are observed in several classes of
slowly rotating A-type stars. For instance, for Am stars inexplicably broad profiles of strong 
line are observed. Large microturbulent velocity is often deduced in 1-D model atmosphere analyses
and line bisectors show blue asymmetry (Landstreet \cite{l98}), which is opposite to the
behaviour of the solar-type stars. In cool magnetic Ap stars macroturbulence of $\ge$\,10~\kms\
is required to reproduce the width of rare-earth lines formed in the upper atmospheric layers
(Kochukhov \& Ryabchikova \cite{kr01}), suggesting that turbulence is not fully suppressed
even in the presence of strong ordered magnetic field. These observations indicate that 
convection is important for the atmospheric dynamics of A-type stars.

\section{Numerical simulations with \cob}

\begin{figure}[!th]
\includegraphics{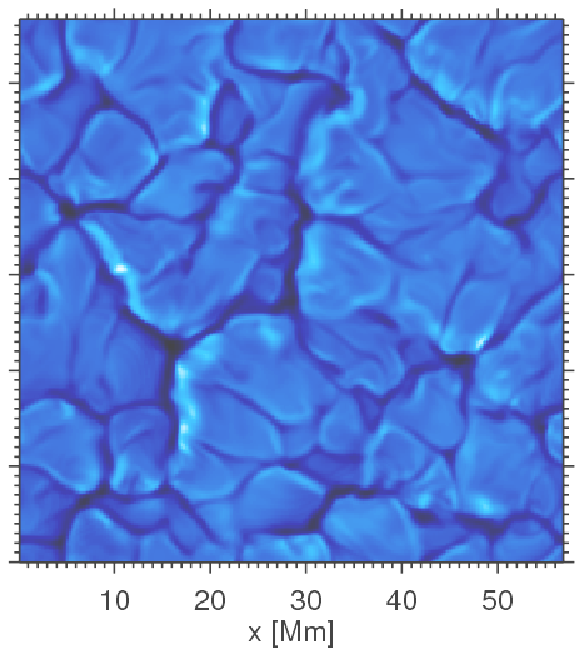}
\includegraphics[width=7.5cm]{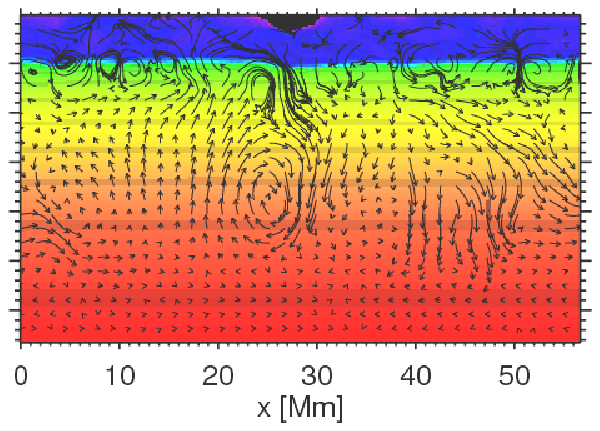}
\caption{\textit{Left panel:} continuum intensity resulting from the 3-D hydrodynamic
simulations of A-star convection. \textit{Right panel:} illustration of the vertical
structure of the 3-D model. Arrows represent velocity field, greyscale plot shows 
temperature distribution.}
\label{fig1}
\end{figure}

3-D hydrodynamic simulations of convection in A stars are considerably more difficult compared
to similar modelling of the solar-type stars. Short radiative time-scales limit the time step
of simulation and thus require 10 to 100 more computing time for an A-type star compared to the
corresponding simulation of the solar-type granulation. Furthermore, simulations need to resolve 
the steep temperature jump below the photosphere and have to cover simultaneously the small surface
granules and the large deep cells.

Here we develop further the 3-D A-star model of Freytag \& Steffen (\cite{fs04}) and present
first results of the new study that has finally overcome computational problems
associated with simulating the A-star convection zone.

We have used the \cob\ code (Freytag \cite{cobman}, Wedemeyer et al. \cite{w04}) to perform 3-D
simulations of the surface convection zone in a main sequence A-type star ($T_{\rm eff}=8000$~K,
$\log g=4.0$). For solving
hydrodynamics equations \cob\ employs conservative finite volume approach.
3-D radiation transport is based on a modified Feautrier scheme using long characteristics.
We use grey or wavelength dependent
opacities derived from ATLAS6 opacity distribution functions. 
Simulations are carried out in Cartesian geometry, assuming periodic lateral boundary
conditions. Top and bottom boundaries are closed for flows and transparent for radiation.
$220\times220\times170$ grid points are used, corresponding to the simulation box with size
$57\times57\times34.5$~Mm.

Fig.~\ref{fig1} illustrates the typical surface granulation pattern and the 
vertical structure of the velocity field deduced with \cob.

\section{Spectrum synthesis and comparison with observations}

\begin{figure}[!th]
\centering
\includegraphics[width=4.8cm,angle=90]{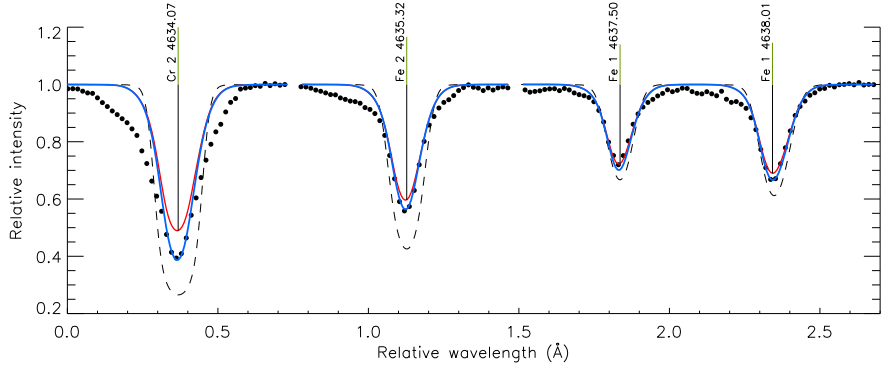}
\caption{Comparison of the observed spectrum of Am star HR~4750 (symbols) with the
theoretical line profiles computed using grey (thin solid curve) and non-grey (thick solid curve) \cob\
models. The dashed line shows 1-D spectrum synthesis using microturbulence $\xi_t=3$~\kms\ and
no macroturbulent or rotational broadening.}
\label{fig2}
\end{figure}

A new 3-D spectrum synthesis code (Piskunov, in preparation) was used to compute line profiles from
the \cob\ snapshots. For A-star models we used the LTE source function and computed line and
continuum  opacities as in 1-D Synth code by Piskunov (\cite{p92}). Radiative transfer is
solved with a short characteristic formal solver based on the quadratic Bezier-spline
approximation of the source function. The flux spectra are produced from the intensity profiles
computed for 29 directions.

We compared our theoretical spectra with the high-resolution observations of the slowly rotating
A2m star HR~4750 ($T_{\rm eff}=8100$~K, $\log g=4.1$, Landstreet \cite{l98}). A set of Fe and
Cr spectral lines in the 4612--4645~\AA\ wavelength region was synthesized, adopting chemical
abundances, stellar and atomic parameters from Landstreet (\cite{l98}).

Fig.~\ref{fig2} presents a comparison of observations and 3-D spectrum synthesis for selected
spectral lines. Our 3-D model reproduces the line cores of strong spectral features, but fails to fit
the observed wide wings and strong blueward asymmetry. On the other hand, weak lines are
successfully reproduced. 
In contrast to the solar-type stars, bisectors computed for our 3-D hydrodynamic models of
A stars are not skewed to the red. In our calculations, the resulting bisectors span a small velocity range and
have a weak blue asymmetry in the outer wings -- a trend which agrees with observations.

\section{Conclusions}

We have carried out 3-D numerical simulations of the A-star convection zone using the radiation
hydrodynamics code \cob. These are the first 3-D RHD calculations for A stars which employ
non-grey opacities. The granulation pattern predicted for A stars appears to be generally similar
to the one found in the solar-type stars, albeit with some distinctive features (bright rims of
granules, large horizontal velocity component). Comparison of the detailed 3-D
spectrum synthesis with \cob\ snapshots and high-resolution observations shows that our 3-D
model successfully reproduces the shape of weak lines in the Am star HR~4750. This is the first
time when a 3-D RHD model is able to fit the A-star line profiles without introducing fudge
line broadening parameters. Theoretical models fail to reproduce the broad wings and prominent
blueward asymmetry observed in the strong spectral lines, indicating that further improvements
(realistic abundances, better resolution) of 3-D models are required.

\begin{acknowledgments}
OK thanks IAU and the Czech Academy of Sciences for supporting his participation
in the 26th IAU General Assembly. This work was also supported by the
travel grant from the Royal Swedish Academy of Sciences.
We thank Prof. J. Landstreet for providing us with the high-resolution observations
of A stars.
\end{acknowledgments}

\end{document}